\documentclass[twocolumn,prx,noshowpacs,superscriptaddress,preprintnumbers,amsmath,amssymb]{revtex4}
\pdfoutput=1
\usepackage{verbatim}
\usepackage{natbib}
\usepackage[dvips]{graphicx}
\usepackage{caption}
\usepackage{xspace}
\usepackage{color}
\usepackage{array}
\usepackage[english]{babel}

\usepackage{multirow}

\newcommand{\bi}{BiSbTeSe$_2$\@\xspace}
\newcommand{\phii}{Physics Institute II, University of Cologne, Z\"{u}lpicher Stra{\ss}e 77, D-50937 K\"{o}ln, Germany}
\newcommand{\phru}{Institute of Metal Physics, S. Kovalevskoy 18, 620219 Ekaterinburg GSP-170, Russia}
\newcommand{\urfu}{Ural Federal University, Mira Street 19, 620002 Ekaterinburg, Russia}

\begin{document}
	\title{Phonon mode calculations and Raman spectroscopy of the bulk-insulating topological insulator BiSbTeSe$_2$}
	\author{Raphael German}
	\affiliation{\phii}
	\author{Evgenia V. Komleva}
	\affiliation{\phru}
	\author{Philipp Stein}
	\affiliation{\phii}
	\author{Vladimir G. Mazurenko}
	\affiliation{\urfu}
	\author{Zhiwei Wang}
	\affiliation{\phii}
	\author{Sergey V. Streltsov}
	\affiliation{\phru}
	\affiliation{\urfu}
	\author{Yoichi Ando}
	\affiliation{\phii}
	\author{Paul H.M. van Loosdrecht}
	\affiliation{\phii}

	\date{\today}
	\hyphenation{Raman}
	
	\begin{abstract}
		The tetradymite compound \bi is one of the most bulk-insulating three-dimensional topological insulators, which makes it important in the topological insulator research. It is a member of the solid-solution system Bi$_{2-x}$Sb$_x$Te$_{3-y}$Se$_y$, for which the local crystal structure, such as the occupation probabilities of each atomic site, is not well understood. We have investigated the temperature and polarization dependent spontaneous Raman scattering in \bi, revealing a much higher number of lattice vibrational modes than predicted by group theoretical considerations for the space group R$\bar{3}$m corresponding to an ideally random solid-solution situation. The density functional calculations of phonon frequencies show a very good agreement with experimental data for parent material Bi$_2$Te$_3$, where no disorder effects were found. In comparison to Bi$_2$Te$_3$ the stacking disorder in \bi causes a discrepancy between theory and experiment. Combined analysis of experimental Raman spectra and DFT calculated phonon spectra for different types of atomic orders showed coexistence of different sequences of layers in the material and that those with Se in the center and a local order of Se-Bi-Se-Sb-Te, are the most favored. 
		
	\end{abstract}
	\maketitle

	\section{Introduction}
	
Three-dimensional (3D) topological insulators (TIs) attract a great deal of interest mainly due to their topologically-protected metallic surface states in which the spin is locked to the momentum \cite{Ando2013,Hasan2010}. Such a spin-momentum locking provides opportunities for realizing various useful functionalities, such as the control of spin polarization for spintronics or the creation of Majorana zero modes for topological quantum computation. To take advantage of the peculiar surface state properties, it is useful to realize the situation that the current flows only through the surface states; this requires that the bulk of a 3D TI should be insulating. In this regard, some of the tetradymite compounds, such as Bi$_2$Se$_3$, Bi$_2$Te$_3$, and Sb$_2$Te$_3$, are prototypical 3D TIs \cite{article} and they have been intensively studied in recent years, but these binary tetradymite compounds are not really insulating in the bulk. 

The first bulk-insulating 3D TI material, Bi$_2$Te$_2$Se, was discovered in 2010 and it is an ordered ternary tetradymite coumpound \cite{PhysRevB.82.241306}. Its ordered nature, with the Se layer conceiled in the middle of the Te-Bi-Se-Bi-Te quintuple-layer (QL) unit, is the key to its bulk-insulating property. To further improve the bulk-insulating property, Ren {\it et al.} have employed the concept of compensation and synthesized a series of solid-solution tetradymite compounds, whose compositions are generally written as Bi$_{2-x}$Sb$_x$Te$_{3-y}$Se$_y$ (BSTS), that are bulk-insulating \cite{Ren2011,Taskin2011}. In such BSTS compounds, the concealment of the Se layer in the middle of the QL unit is kept intact, but an appropriate mixing in the occupancy of the anion layers as well as the outer chalcogen layers makes it possible to achieve a high level of compensation between residual electron- and hole-doping. In particular, the nominal composition of BiSbTeSe$_2$ (BSTS2), which corresponds to ($x$,$y$) = (1,2) variant of the BSTS solid solution, was found to achive a particularly high level of bulk-insulation \cite{Ren2011,Arakane2012,Segawa2012}. Since BSTS2 is currently the most bulk-insulating 3D TI material available as a bulk single crystal, it is widely used in various experiments aiming at studying the surface transport properties \cite{QHE-Xu2014, Yang2016, Ghatak2018, Jauregui2018, Lu2018}. 
Despite the importance of BSTS2 in the research of 3D TIs, little is know about its local crystal structure. For example, the extent of randomness in the Bi/Sb sites and the outer Te/Se sites in the QL layers has not been studied. If there is some correlations between the occupancies of these nominally random sites, it may lead to a local inversion-symmetry breaking, which might affect/limit the topological properties of this compound. In this respect, the high-frequency phonon modes detected in Raman scattering are sensitive to local symmetry of the crystal lattice, and hence they are expected to provide useful information on the local crystal structure of BSTS2.

In the present paper, we report combined experimental (Raman) and theoretical (density functional theory - DFT) study of BiSbTeSe$_2$ solid solution, which allowed us to obtain insights into the local structure of BiSbTeSe$_2$. The temperature- and polarization-dependent Raman spectra show the existence of an unexpectedly high number of Raman active modes, which are forbidden in the group-theoretical considerations of the global symmetry of BSTS2 ($R\bar{3}m$ \cite{Efthimiopoulos2013}), signaling a high degree of local symmetry-breaking disorder. To understand their origin, we performed DFT calculations with different combinations of atomic layers in the unit cell. This analysis points to the conclusion that the occupancy of the upper and lower sides of a QL unit is likely to be correlated to prefer the local structure of Se-Bi-Se-Sb-Te or Te-Sb-Se-Bi-Se, which breaks inversion symmetry. Moreover, we suggest the peak at $\sim$120 cm$^{-1}$ in the Raman spectra corresponding to the $E_g$ phonon can be potentially used as a marker to estimate the degree of this local inversion symmetry breaking.

	\begin{figure}[t!]
		\centering
		\captionsetup{justification=raggedright,singlelinecheck=false}
		\includegraphics[width=.45\textwidth]{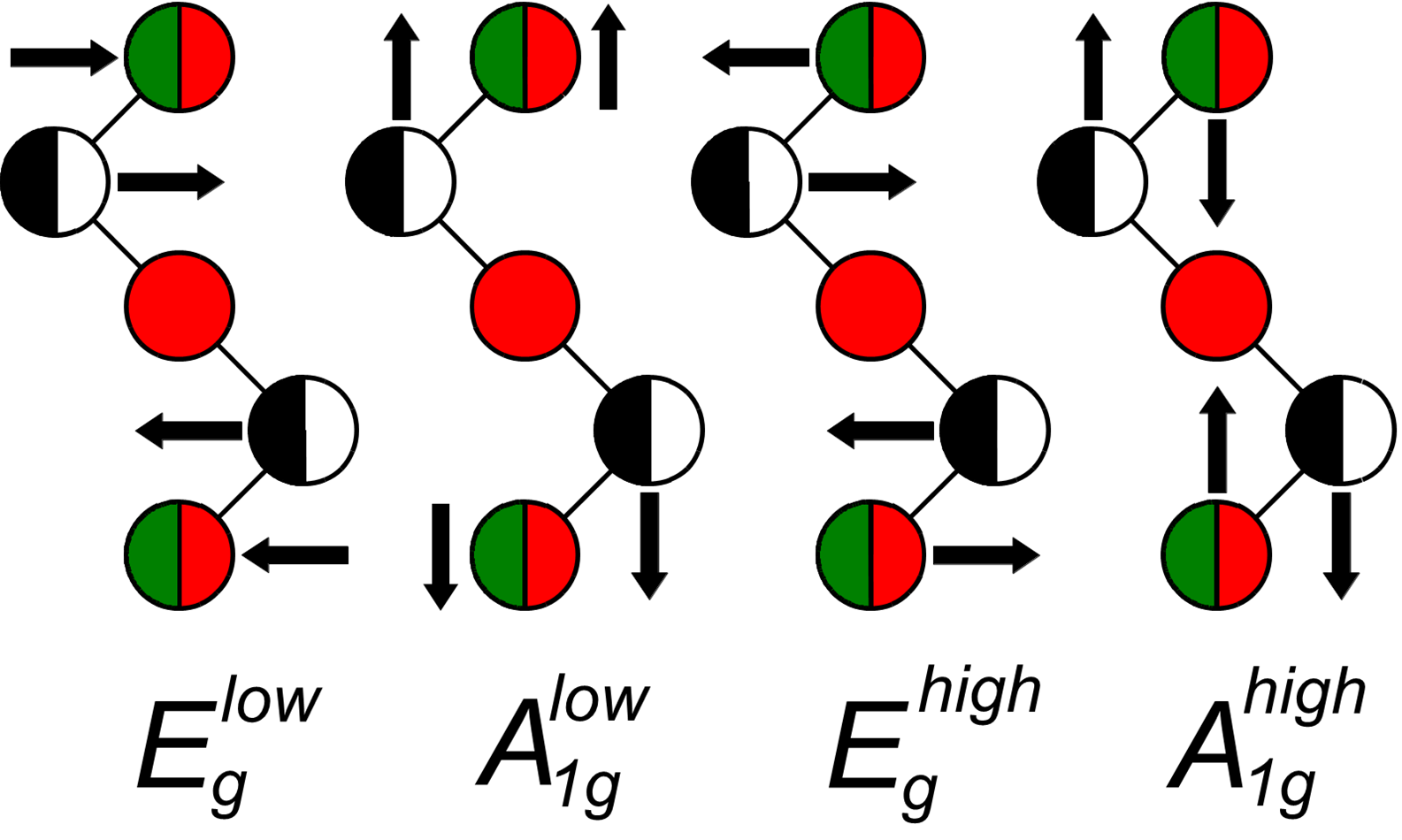}
		\caption{Displacement patterns of the ions, Te (green), Se (red), Bi and Sb (either black or white), in \bi across the quintuple-layered unit cell, which generate Raman active modes.  Low and high refers to the frequency.}
		\label{modes}
	\end{figure}
	
	\section{Experimental details}
	The single crystals of \bi used in the present study were grown from high-purity elements by using a modified Bridgman method in a sealed quartz glass tube.\cite{Ren2011}
	Before the measurement, the crystals surface was cleaved along the (001) plane for a clean shiny surface.
	For the spontaneous Raman measurements a continuous wave DPSS laser of 532 nm wavelength and a Krypton gas laser of 647.1 nm wavelength were used. The samples were mounted in an Oxford Instruments Microstat cryostat, cooled by liquid Helium. 
	Raman spectra were collected using a triple grating spectrometer operating in subtractive mode equipped with a LN$_2$ cooled CCD camera.  
	
	\section{Results}
	\subsection{Raman spectroscopy}
	\begin{figure*}[t!]
		\includegraphics[width=.45\textwidth]{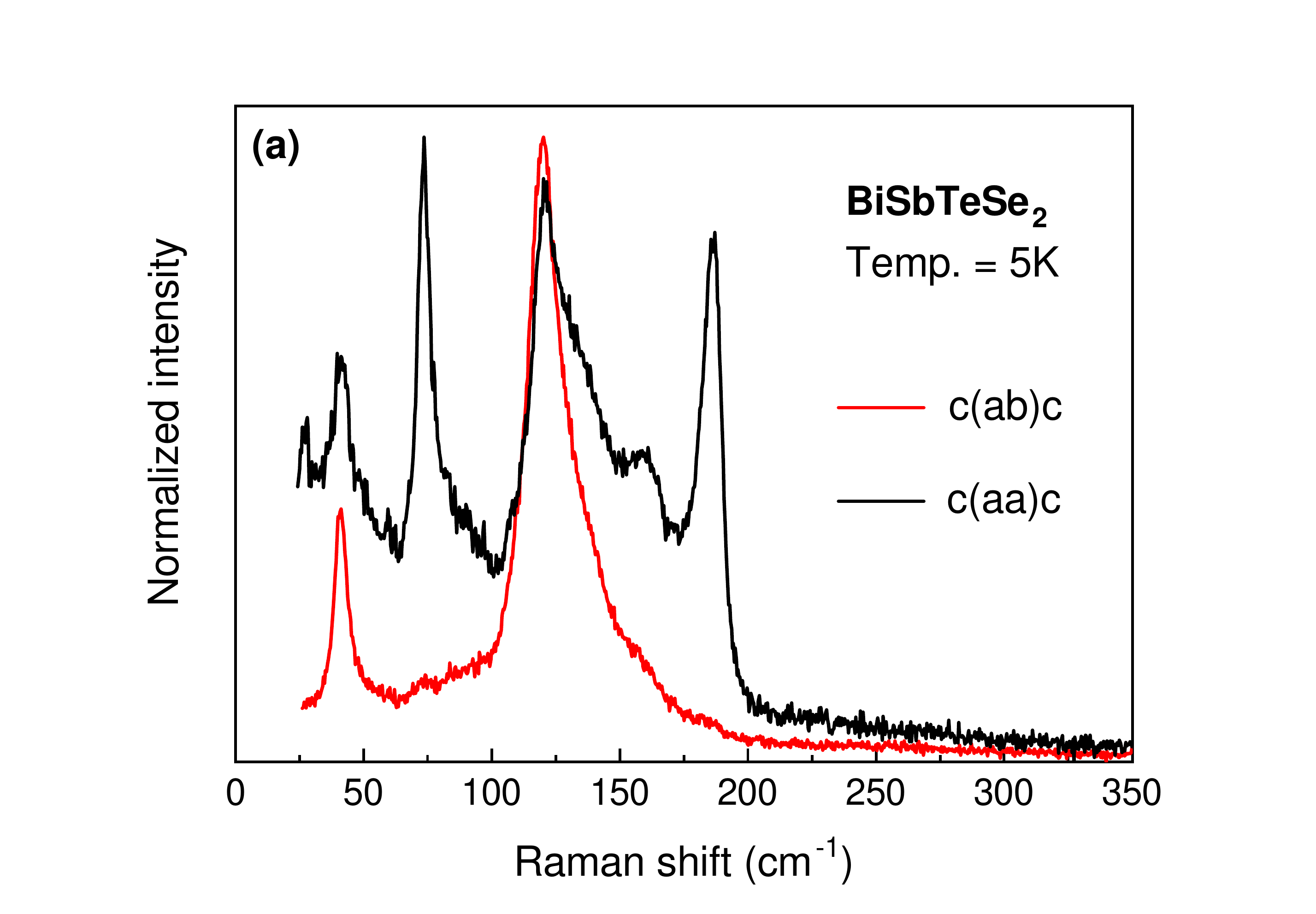} 
	    \includegraphics[width=.4\textwidth]{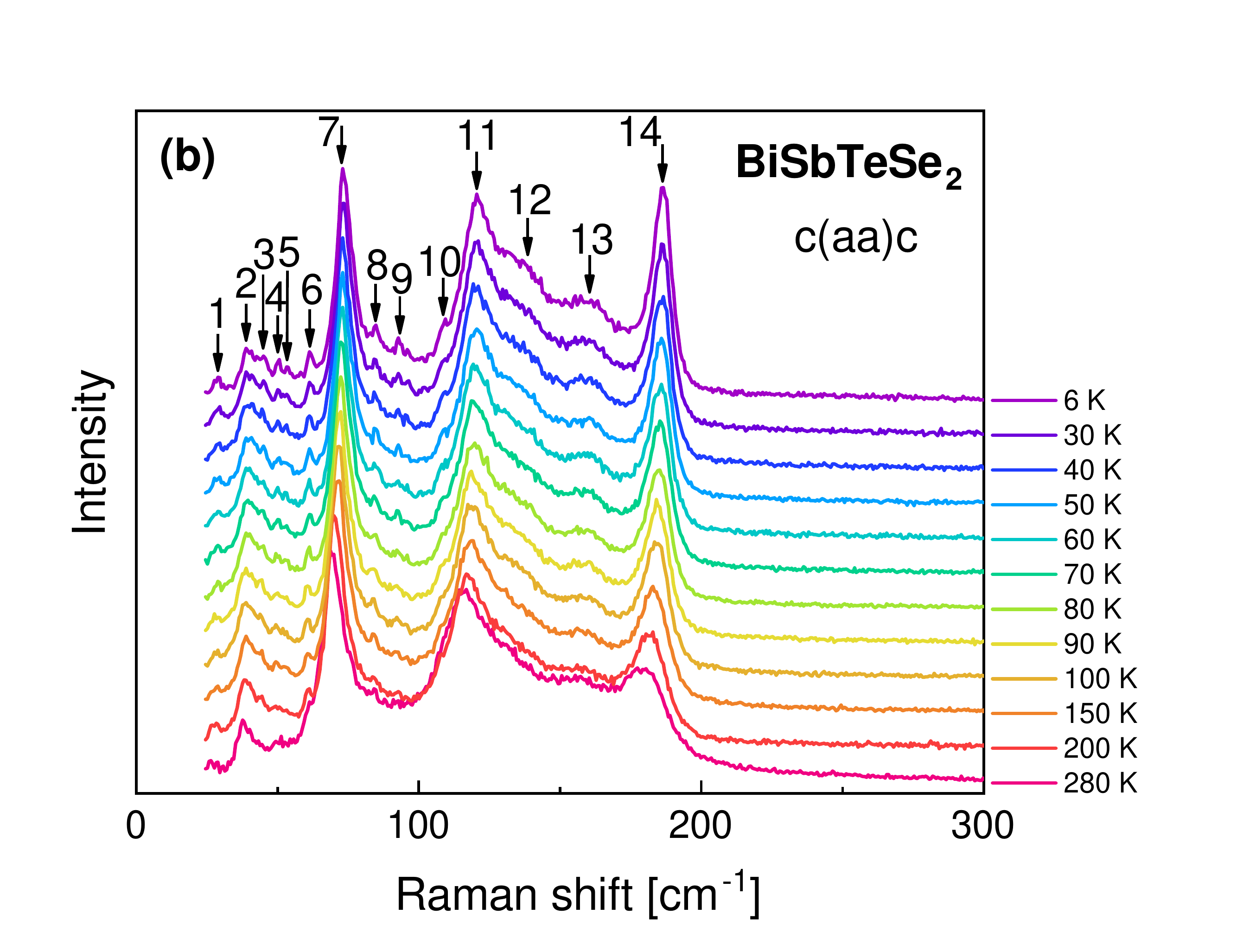} 
	\captionsetup{justification=raggedright,singlelinecheck=false}
		\caption{(a) Normalized Raman spectra of \bi at 5K in c(aa)c and c(ab)c geometry, in black and red, respectively. (b) Temperature dependent Raman spectra of BiSbTeSe$_2$, measured in c(aa)c. Arrows denote the positions of the fitted oscillators, see Table \ref{table}.} 
		\label{pol_dependence}
	\end{figure*}
	
	Assuming the same primitive unit cell for \bi as for Bi$_2$Te$_3$, Bi$_2$Se$_3$,  and Sb$_2$Te$_3$, five atoms are decisive for the group theoretical considerations of the vibrational modes. Therefore, in accordance with the chemical formula there are $3N - 3 = 12$ optical lattice dynamical modes at q = 0.
	These modes are exclusively Raman- or infrared-active according to inversion symmetry and consequently to their selection rules \cite{Richter1977}. If \bi is a ideally random solid solution, the global symmetry of R$\bar{3}$m might be kept as for the parent compounds and then the irreducible representation of the Raman active modes is given by
	\begin{equation}
	\Gamma_{vib} = 2A_{1g}+2E_{g}, 
	\label{irreps} 
	\end{equation} 
	where $E_{g}$ and $A_{1g}$ denote in-plane and out-of-plane vibrational modes, see Fig. \ref{modes}. Locally, this statement does not necessarily hold anymore. Inversion symmetry automatically breaks when the QL are not symmetrically occupied by the same atoms. Thus, the originally assumed global space group of $R\bar{3}m$ for \bi turns locally into $R3m$ with the following irreducible representation:
	\begin{eqnarray}
	\Gamma_{vib} = 4A_1 + 4E.
	\label{irreps2}
	\end{eqnarray}
	Parallel and cross polarizations, i.e. c(aa)c and c(ab)c in Porto notation, probe the diagonal and off-diagonal elements of the Raman tensor, respectively.
	Thus, E modes are observable in both c(aa)c and c(ab)c geometry, while A$_1$ are only visible in c(aa)c geometry. In the following, we use this more general representation to describe our results.
    Fig. \ref{pol_dependence}a shows the polarization dependent Raman measurements of \bi on the (001)- surface at 5 K. The excitation wavelength of the laser was 647.1 nm and all measured samples were freshly cleaved before measurements along the c- direction.
	Two distinct modes ($\sim$ 40 cm$^{-1}$ and $\sim$ 120 cm$^{-1}$)  appear in both polarizations,  which are associated with the doubly-degenerate in-plane vibrational $E$ modes. This is line with the number of reported $E_{g}$ modes for the binary parent compounds, listed together with our fit results in Table \ref{table}. 
	Consequently, the remaining modes are assumed to correspond to the out of plane vibrational $A_{1}$ modes, which exceed the number given by the group theoretical considerations. 
	A couple of vibrational energies, reported for similar compounds, are visible in \bi too, see Table \ref{table}. Notably, the modes of Bi$_2$Se$_3$, Bi$_2$Te$_2$Se , and  Sb$_2$Te$_2$Se correspond very well to the energies of the most distinct peaks 7, 11, 12, 13, and 14, respectively, leading to the assumption that the configuration with the Selenium layer in the middle of the QL is indeed the most favoured one and that the compound is highly mixed. In this sense, the obtained Raman spectrum is a superposition of locally differing sequences of ions.
		 
\begin{figure*}[tbh]
	\includegraphics[width=.3\textwidth]{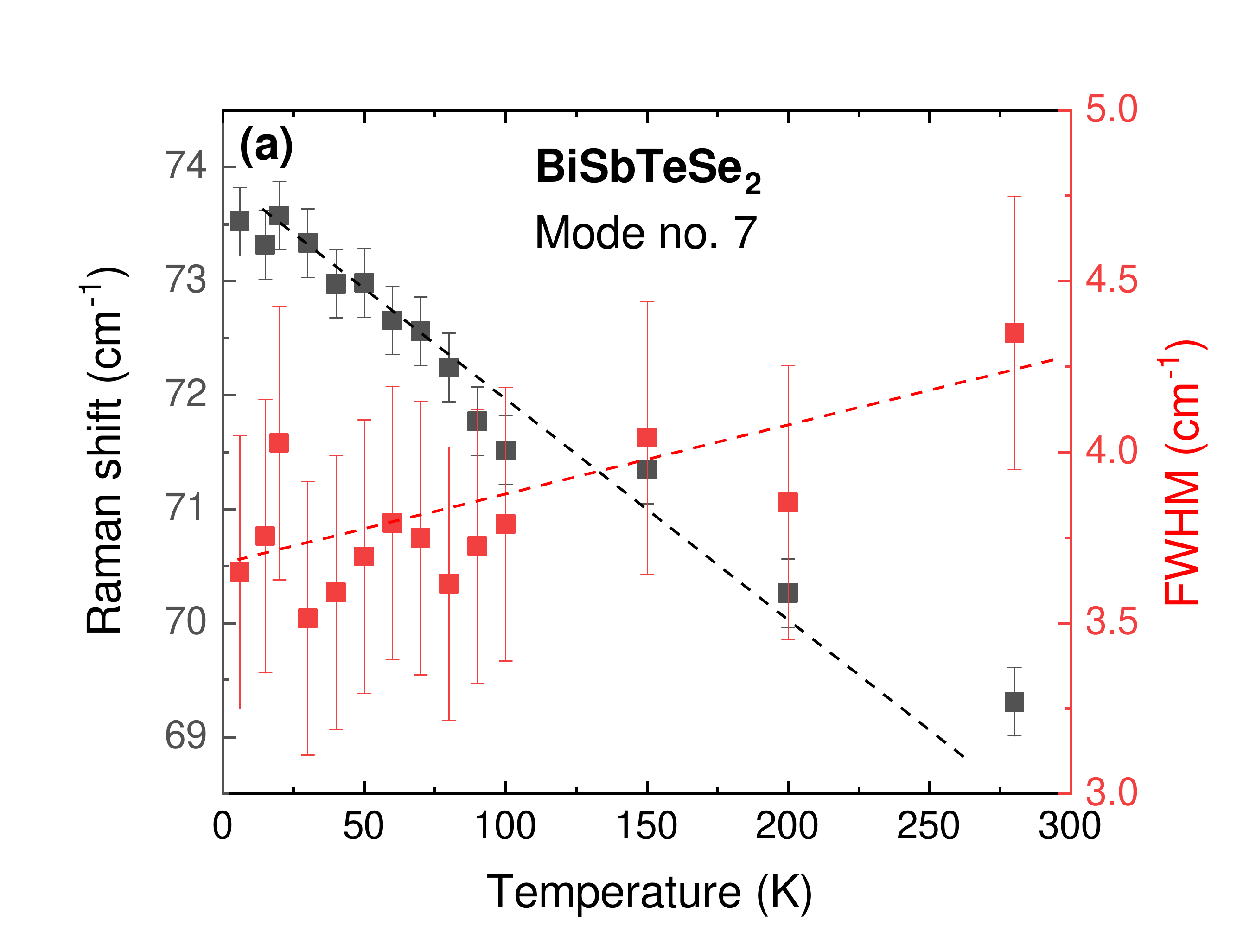}\vspace{1mm}  
	\includegraphics[width=.3\textwidth]{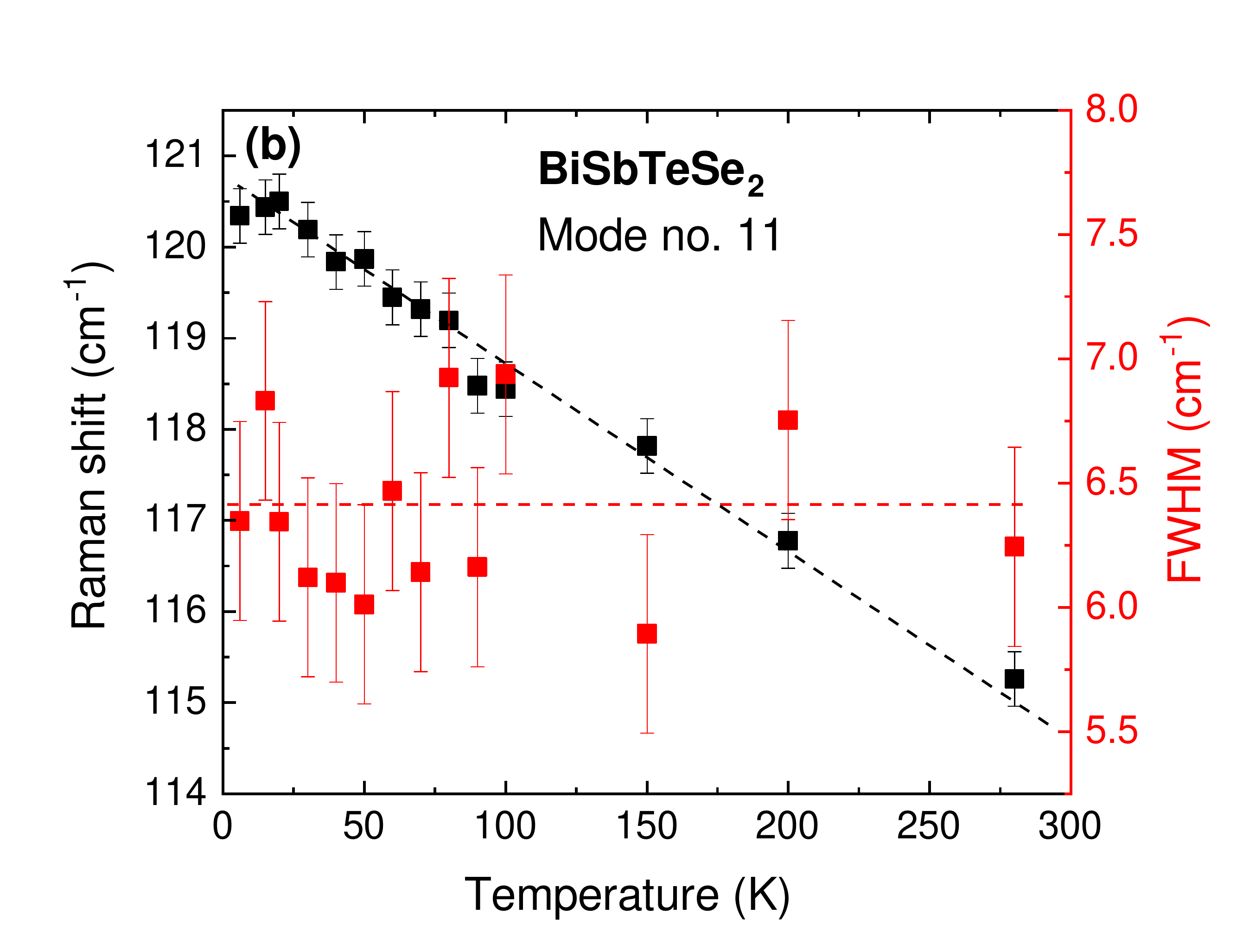}\vspace{1mm} \includegraphics[width=.3\textwidth]{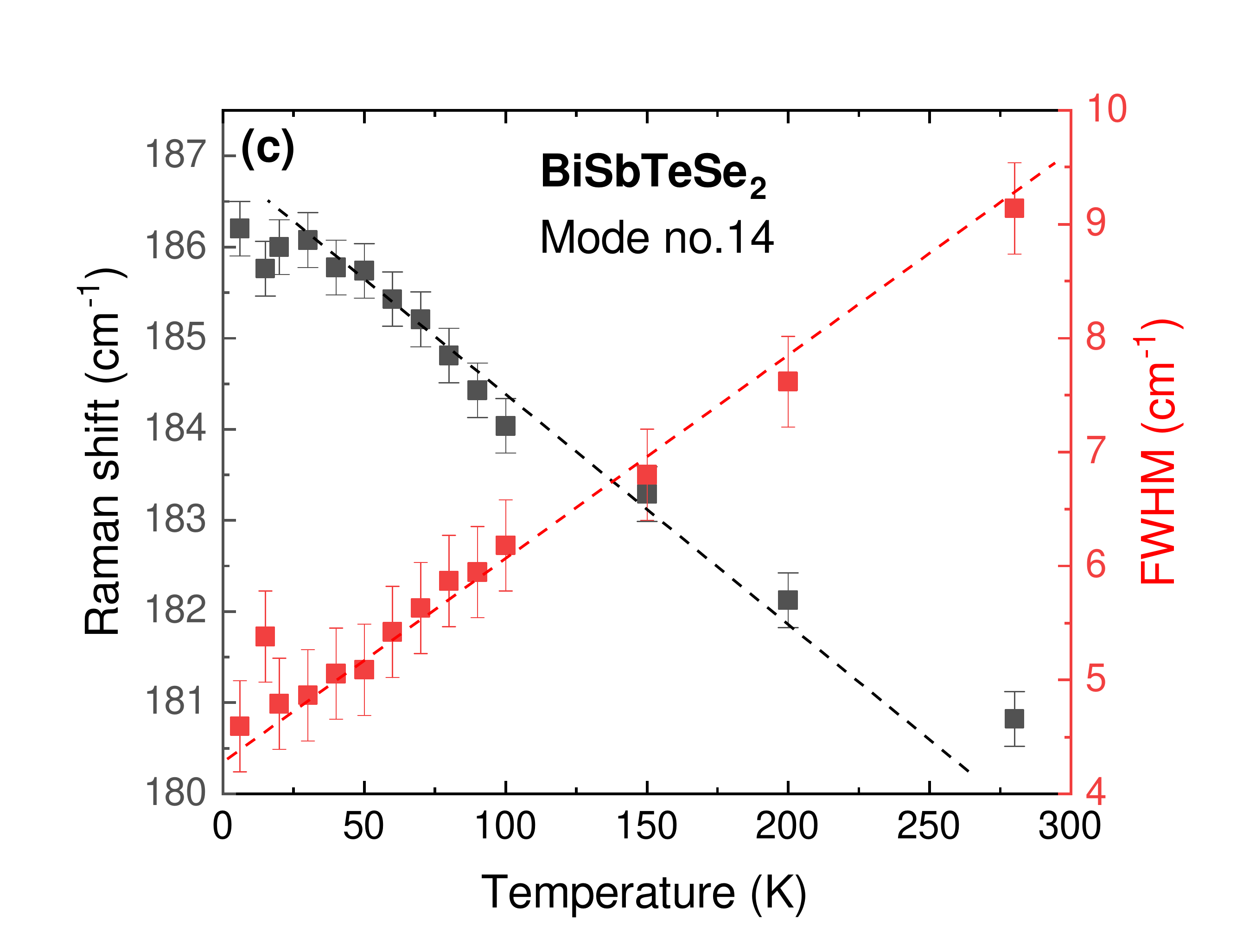}\vspace{1mm} 
	\captionsetup{justification=raggedright,singlelinecheck=false}
	\caption{Temperature dependent energy shifts and FWHM of (a) mode no. 7 (b) no. 11 (c) no. 14. Dotted lines are guides for the eyes.}
		\label{energy_shift}
    \end{figure*}

	To shed more light on this, we performed temperature dependent Raman measurements, presented in Fig. \ref{pol_dependence}b. These spectra were measured from 280 K down to 6 K and each spectrum was normalized by the total spectral weight (SW), which decreases as expected by the Bose factor. The temperature dependent measurements were done with a laser excitation wavelength of 532 nm. Compared to the laser excitation wavelength of 647.1 nm, no differences, in terms of energy shifts or line widths of the modes, were observed.
	Numbered arrows depict the central positions of fitted oscillators.
	It was suggested that electron-phonon coupling could play a role in this class of compounds \cite{Akrap2012}, revealed by possible asymmetric Fano line shapes. In \bi, the most likely mode for showing this interaction is no. 14, which shows a strong deviation from a pure Lorentzian line shape, especially below 80 K. However, according to the high number of phonon modes, it is not possible to make consistent assumptions about the underlying continuum. Therefore, we cannot confirm a Fano-like resonance behaviour with our data. The best fitting results were achieved by applying a Pseudo-Voigt-function, which reflects usually the instrumental limitation in the case of a very sharp mode. 
	Fig. \ref{energy_shift} shows the temperature dependencies of the Raman shifts and full width at half maximums (FWHM) for the strongest modes no. 7, 11, and 14. They all show anharmonic phonon decaying behaviour. For the mode no. 14, this effect is more pronounced, which implies stronger anharmonic coupling to the lattice. Below $\sim 40$ K the observed variations with temperature saturate. This coincides with the reported temperature of charge puddle formation \cite{Borgwardt2016}. 
	
	\begin{table*}[t]
		\begin{tabular}{c c c c c c c c c c}  
		\hline 	\hline
			\multicolumn{4}{ c }{BiSbTeSe$_2$} & \multicolumn{2}{c}{Bi$_2$Te$_3$}& Bi$_2$Te$_2$Se &Sb$_2$Te$_2$Se & Bi$_2$Se$_3$ & Sb$_2$Te$_3$\\  
			No. & Exp. & FWHM & Exp. Mode& Exp.& Cal.& Exp.& Exp.& Exp.& Exp.\\
			\hline
			1  & 28.5 			& 2.8 			&	& 25 &		& 		 & 		 & & \\
			2 	 & 39.3 			& 4.6 	& E		& 35.8 (E$_{g}$) 	& 39.1 (E$_{g}$)	& 		 & 41.1 (E$_{g}$)	 & & \\
			3 	 & 44.4 			& 5.9 	& A$_{1}$		&	 	 		& 		&	 	& 		 & & \\
			4 	 & 50.3 			& 2.4 	& A$_{1}$		& 	 			& 		&	 	& 		 & & \\
			5 	 & 53.5 			& 3.0 	& A$_{1}$		& 	 			& 60.3 (E$_{u}$)		&	 	& 		 & & \\
			6 	 & 61.4 			& 1.7 & A$_{1}$			& 62.5 (A$_{1g}$) & 63.6 (A$_{1g}$)	&	 	& 		 & & \\
			7 	 & 73.6			& 6.9 	& A$_{1}$		& 				& 		& 67.7	(A$_{1g}$)& 		 & 72 (A$_{1g}$) & 69 (A$_{1g}$)\\
			8 	 & 84.7 			& 9.1 	& A$_{1}$		& 				& 94.8 (E$_{u}$)  			& 		& 82.2 (A$_{1g}$)	 &  &\\
			9 	 & 93.3 			& 5.3  & A$_{1}$			&				& 99.3 (A$_{1u}$)			& 		& 		 &  &\\
			10  & 108.5 		& 2.1 	& A$_{1}$		& 103 (E$_{g}$)	& 103.2 (E$_{g}$)	& 110.7 (E$_{g}$)	& 		 &  & 112 (E$_{g}$)\\
			11  & 120.0 		& 6	& E		& 				& 125.0 (A$_{1u}$)		&	 	& 117.5 (E$_{g}$) & 121 ( E$_{g}$) &\\
			12  & 135.2 	& 12  & A$_{1}$				& 134 (A$_{1g}$)	& 136.1 (A$_{1g}$)	& 142	 & 	138.3 (E$^L$)	 &  &\\
			13  & 161.8 	& 20	& A$_{1}$			&				& 			& 157.7	(A$_{1g}$) & 		 &  & 165 (A$_{1g}$)\\
		
			14  & 186.6 	& 4.6 	& A$_{1}$			& 				&		& 		 & 188.9 (A$_{1g}$) & 174.5 (A$_{1g}$) &\\
			\hline 	\hline
		\end{tabular}
		\captionsetup{justification=raggedright,singlelinecheck=false}
		\caption{Fitted Raman shifts and FWHM in cm$^{-1}$ for \bi at 5K, with experimental mode assignments, concluded from polarization measurements.} The Raman shifts are compared to similar compounds with their mode assignments (in brackets), given from the literature for Bi$_2$Te$_2$Se \cite{Akrap2012,Tian2016}, Sb$_2$Te$_2$Se \cite{Lee}, Bi$_2$Se$_3$ \cite{Richter1977}, Sb$_2$Te$_3$ \cite{Richter1977}. E$^L$ denotes a local mode, claimed by the others of Ref. \cite{Lee}. For Bi$_2$Te$_3$ the calculated values are given, too.
		\label{table}
	\end{table*}

	\subsection{Density functional calculations of phonon modes}
	To investigate the possible origins of the observed vibrational modes, DFT calculations of lattice dynamics were performed. First we calculated the well investigated parent compound Bi$_2$Te$_3$ and found a rather good agreement of phonon frequencies with the experimental results (see Table \ref{table}). Maximal deviation between theoretical and experimental frequencies does not exceed 3.3 cm$^{-1}$. 
	
\begin{figure}[htbp]
		\centering
		\includegraphics[width=.45\textwidth]{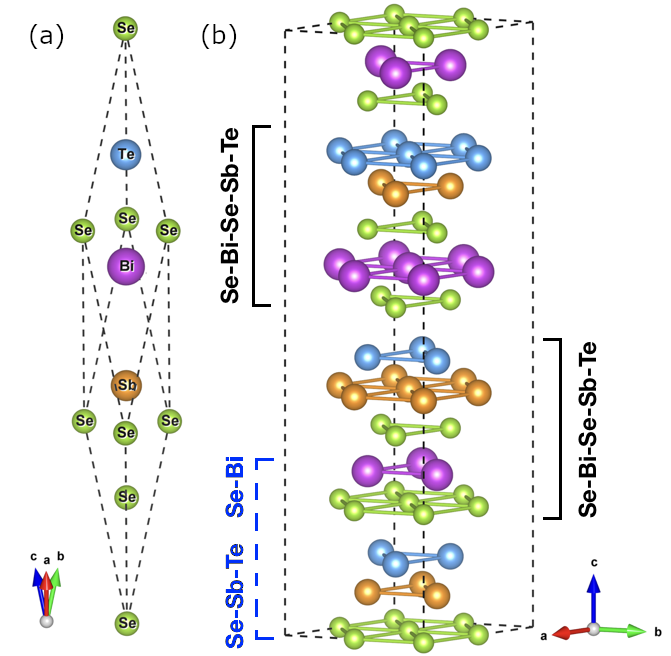}
		\captionsetup{justification=raggedright,singlelinecheck=false}
		\caption{Crystal structure of \bi in R3m with (a) rhombohedral and (b) hexagonal unit cells.}
		\label{structure}
	\end{figure}

	The binary compound lattice parameters and atomic positions were taken from literature.\cite{hulliger_lévy} The rhombohedral axes for \bi are $a = 10.09326$ $\mathring{A}$ and $ \alpha = 23.7852^\circ$ \cite{Ren2011}. As the sequence of atomic layers in \bi is unknown and, as it was mentioned above, one would expect a stacking disorder,  we arranged atoms in the unit cell in three different ways: Se(1)-Sb-Se(2)-Bi-Te, Se(1)-Bi-Se(2)-Sb-Te and Se(1)-Sb-Te-Bi-Se(2). Previously equivalent atomic positions are now occupied by different sorts of atoms, breaking the inversion symmetry.
	Resulting structure for \bi and the unit cell used in the calculations are shown in Fig.\ref{structure}.
	
	We used the frozen-phonon method \cite{phonopy} and density functional theory (DFT) to calculate phonon frequencies. The exchange-correlation functional was chosen to be in the form proposed by Perdew et $al.$ \cite{PhysRevLett.77.3865,PhysRevLett.78.1396}. Since the spin-orbit coupling (SOC) was shown to be important in the formation of topologically protected states in the parent compounds Bi$_2$Te$_3$ and Bi$_2$Se$_3$ \cite{PhysRevB.61.8162,article}, we included this interaction in the calculation scheme. The calculations were performed using the Vienna ab initio simulation package (VASP) \cite{PhysRevB.47.558,PhysRevB.49.14251,KRESSE199615,PhysRevB.54.11169}. A plane-wave cutoff energy was set to 500 eV. We found that the k-mesh considerably affects the results of the calculations. Present results were obtained with the $8\times8\times8$ k-mesh.
	The structure was geometrically optimized until the energy difference between two ionic iterations reached 10$^{-5}$ eV/atom. To calculate the Hessian matrix only symmetry inequivalent displacements were considered within finite differences method.
	
	In the present calculations for \bi, the three atomic arrangements mentioned above were considered. According to the electronic structure calculations the most favorable sequence is Se(1)-Bi-Se(2)-Sb-Te. 
	Its total energy is lower by 68 meV/f.u. than for Se(1)-Sb-Se(2)-Bi-Te, and lower by approximately 100 meV/f.u. than for Se(1)-Bi-Te-Sb-Se(2). Keeping the same relative order of layers but choosing quintuple layer (QL) unit in another way (e.g. see Fig.\ref{structure}, blue dashed bracket) we got very similar results. However, the sequence with Selenium in the middle of the "real" QL unit still has the lowest total energy.	Thus, from the theoretical point of view the most favoured stacking sequence is Se(1)-Bi-Se(2)-Sb-Te.
	
	The calculation results for Bi$_2$Te$_3$ are listed in Table \ref{table} and for different sequences of atomic layers in \bi in Table \ref{table2}. The mode assignment was done using the analysis of the eigenvectors of the obtained dynamical matrix. The atomic displacements for a given frequency are in good agreement with the group theory consideration. As has been mentioned above, the calculated frequencies for Bi$_2$Te$_3$ show a good agreement with the experiment and lead to the conclusion that DFT should work for \bi, too. Since there are too many $A_1$ modes due to disorder and it is nearly impossible to make a correspondence between calculated and experimental frequencies, we compare $E$ phonons. These lines were clearly identified in the polarization measurements.
	
	In Table II, one can see that the low frequency $E$ line is found in all three possible structural stackings. In contrast, the observed $E$ mode at 120 cm$^{-1}$ fits only with the calculated $E$ mode of Se(1)-Bi-Se(2)-Sb-Te sequence. Thus, this analysis suggests that locally our samples mostly have Se(1)-Bi-Se(2)-Sb-Te order. The conclusion that the stacking order is mostly Se(1)-Bi-Se(2)-Sb-Te also agrees with the DFT total energy calculations presented above.

		\begin{table*}[t]
		\begin{tabular}{c c c c c c}
			\hline
			\hline
			Mode assignment & Se-Bi-Se-Sb-Te & Se-Sb-Se-Bi-Te & Se-Sb-Te-Se-Bi & Se-Bi-Te-Se-Sb & Te-Sb-Se-Se-Bi \\ 
			\hline
			A$_1$ & 180.0 & 176.1 & 180.0 & 175.9 & 178.8  \\ 
			A$_1$ & 158.1 & 150.8 & 157.9 & 150.6 & 159.0 \\ 
			A$_1$ & 133.8 & 138.0 & 133.7 & 137.7 & 117.3 \\ 
			E & 128.0 & 129.7 & 127.8 & 129.5 & 131.9 \\ 
			E & 119.6 & 109.8 & 119.7 & 109.7 & 109.4 \\
			E & 82.1 & 74.3 & 82.1 & 74.1 & 68.9 \\ 
			A$_1$ & 70.3 & 69.2 & 70.5 & 69.0 & 74.2 \\ 
			E & 42.2 & 42.9 & 42.1 & 42.6 & 45.6 \\ 
			\hline
			Total energy & 0 & 68.1 & 16.6 & 117.6 & 240.0 \\
			\hline
			\hline
		\end{tabular} 
		\captionsetup{justification=raggedright,singlelinecheck=false}
		\caption{Calculated phonon frequencies in cm$^{-1}$ and relative values of total energy in meV for different sequences of atomic layers in \bi.}
		\label{table2}
	\end{table*}

	\section{Discussion}
	
	The DFT calculations of Bi$_2$Te$_3$ reproduce very well the experimentally found phonon energies and symmetries at 5 K, confirming the group theoretical assumption for the space group R$\bar{3}$m. This consideration can, however, hardly be directly converted to \bi since in that case the inversion symmetry is necessarily broken. In this regard, the irreducible representations for the reduced symmetry R3m show an equal number of optical A$_1$ and E modes, see Eq. \ref{irreps2}, and this prediction still does not agree with the experiment, which shows that at least 14 phonon modes are active in \bi. It is useful to note that, in contrast to Eq. 2, there are many more A$_1$ modes than E modes (we only identify mode No. 2 and No. 11 to be phonons in E symmetry), which is puzzling.
	Probably, the high number of $A_1$ phonon modes is due to the superposition of many different atomic sequences, which change locally in the layered structure of \bi. 

	It is useful to note that in Bi$_2$Te$_2$Se or Bi$_2$Te$_{1-x}$Se$_x$ for $0.2 <$ x $<0.9$, some peculiar modes which have neither A$_1$ nor E symmetry have been observed by Tian et al. \cite{Tian2016} and Richer et al. \cite{Richter1977}. These authors called them local modes, which were proposed to result from Se/Te antisite point defects and vibrate at their own frequency with a broad line width. In view of the work, these local modes may actually be related to the local inversion-symmetry breaking due to antisite defects.

	By comparing the phonon energies of the strong modes in the Raman spectra in Fig. 2 to the calculated phonon energies for various local structures, one notices that the sequences where Selenium is in the middle of the quintuple gives the strongest contributions to the phonon spectrum, meaning that in a major portion of the samples, Selenium is sitting in the middle of the quintuple layer.
	This is in line with our DFT calculations, which show that the sequence Se(1)-Bi-Se(2)-Sb-Te has the lowest total energy. 

	\section{Summary}

	We have performed polarization dependent spontaneous Raman scattering in a temperature range from 6 K to room temperature and observed an unexpectedly high number of Raman active modes,  which contradicts the group theoretical considerations for the global symmetry and points to the significance of local inversion symmetry breaking. Our DFT calculations, which explain the Raman spectra of Bi$_2$Te$_3$ very well, help us to identify the most prominent local structure realized in \bi, which turned out to  be the Se(1)-Bi-Se(2)-Sb-Te sequence.

	\section{Acknowledgements}
	
	This work was supported by the Deutsche Forschungsgemeinschaft (DFG) 
	through the Collaborative Research Center SFB 1238 (projects A04 and B03) and by Ural branch of RAS (grant 18-10-2-37), the FASO through research program "spin" AAAA-A18-118020290104-2, Russian ministry of education and science via contract 02.A03.21.0006, the Russian foundation for basic research (RFBR) via project RFBR 16-32-60070 and the Russian president council on science, program MD-
    916.2017.2.
	
	\newpage
	\bibliography{BSTS}

\begin{thebibliography}{28}
\expandafter\ifx\csname natexlab\endcsname\relax\def\natexlab#1{#1}\fi
\expandafter\ifx\csname bibnamefont\endcsname\relax
  \def\bibnamefont#1{#1}\fi
\expandafter\ifx\csname bibfnamefont\endcsname\relax
  \def\bibfnamefont#1{#1}\fi
\expandafter\ifx\csname citenamefont\endcsname\relax
  \def\citenamefont#1{#1}\fi
\expandafter\ifx\csname url\endcsname\relax
  \def\url#1{\texttt{#1}}\fi
\expandafter\ifx\csname urlprefix\endcsname\relax\def\urlprefix{URL }\fi
\providecommand{\bibinfo}[2]{#2}
\providecommand{\eprint}[2][]{\url{#2}}

\bibitem[{\citenamefont{Ando}(2013)}]{Ando2013}
\bibinfo{author}{\bibfnamefont{Y.}~\bibnamefont{Ando}}, \bibinfo{journal}{J.
  Phys. Soc. Japan} \textbf{\bibinfo{volume}{82}}, \bibinfo{pages}{102001}
  (\bibinfo{year}{2013}).

\bibitem[{\citenamefont{Hasan and Kane}(2010)}]{Hasan2010}
\bibinfo{author}{\bibfnamefont{M.~Z.} \bibnamefont{Hasan}} \bibnamefont{and}
  \bibinfo{author}{\bibfnamefont{C.~L.} \bibnamefont{Kane}},
  \bibinfo{journal}{Rev. Mod. Phys.} \textbf{\bibinfo{volume}{82}},
  \bibinfo{pages}{3045} (\bibinfo{year}{2010}).

\bibitem[{\citenamefont{Zhang et~al.}(2009)\citenamefont{Zhang, Liu, Qi, Dai,
  Fang, and Zhang}}]{article}
\bibinfo{author}{\bibfnamefont{H.}~\bibnamefont{Zhang}},
  \bibinfo{author}{\bibfnamefont{C.-X.} \bibnamefont{Liu}},
  \bibinfo{author}{\bibfnamefont{X.-L.} \bibnamefont{Qi}},
  \bibinfo{author}{\bibfnamefont{X.}~\bibnamefont{Dai}},
  \bibinfo{author}{\bibfnamefont{Z.}~\bibnamefont{Fang}}, \bibnamefont{and}
  \bibinfo{author}{\bibfnamefont{S.-C.} \bibnamefont{Zhang}},
  \bibinfo{journal}{Nature Phys.} \textbf{\bibinfo{volume}{5}},
  \bibinfo{pages}{438} (\bibinfo{year}{2009}).

\bibitem[{\citenamefont{Ren et~al.}(2010)\citenamefont{Ren, Taskin, Sasaki,
  Segawa, and Ando}}]{PhysRevB.82.241306}
\bibinfo{author}{\bibfnamefont{Z.}~\bibnamefont{Ren}},
  \bibinfo{author}{\bibfnamefont{A.~A.} \bibnamefont{Taskin}},
  \bibinfo{author}{\bibfnamefont{S.}~\bibnamefont{Sasaki}},
  \bibinfo{author}{\bibfnamefont{K.}~\bibnamefont{Segawa}}, \bibnamefont{and}
  \bibinfo{author}{\bibfnamefont{Y.}~\bibnamefont{Ando}},
  \bibinfo{journal}{Phys. Rev. B} \textbf{\bibinfo{volume}{82}},
  \bibinfo{pages}{241306} (\bibinfo{year}{2010}).

\bibitem[{\citenamefont{Ren et~al.}(2011)\citenamefont{Ren, Taskin, Sasaki,
  Segawa, and Ando}}]{Ren2011}
\bibinfo{author}{\bibfnamefont{Z.}~\bibnamefont{Ren}},
  \bibinfo{author}{\bibfnamefont{A.~A.} \bibnamefont{Taskin}},
  \bibinfo{author}{\bibfnamefont{S.}~\bibnamefont{Sasaki}},
  \bibinfo{author}{\bibfnamefont{K.}~\bibnamefont{Segawa}}, \bibnamefont{and}
  \bibinfo{author}{\bibfnamefont{Y.}~\bibnamefont{Ando}},
  \bibinfo{journal}{Phys. Rev. B} \textbf{\bibinfo{volume}{84}},
  \bibinfo{pages}{165311} (\bibinfo{year}{2011}).

\bibitem[{\citenamefont{Taskin and Ando}(2011)}]{Taskin2011}
\bibinfo{author}{\bibfnamefont{A.~A.} \bibnamefont{Taskin}} \bibnamefont{and}
  \bibinfo{author}{\bibfnamefont{Y.}~\bibnamefont{Ando}},
  \bibinfo{journal}{Phys. Rev. B} \textbf{\bibinfo{volume}{84}},
  \bibinfo{pages}{035301} (\bibinfo{year}{2011}).

\bibitem[{\citenamefont{Arakane et~al.}(2012)\citenamefont{Arakane, Sato,
  Souma, Kosaka, Nakayama, Komatsu, Takahashi, Ren, Segawa, and
  Ando}}]{Arakane2012}
\bibinfo{author}{\bibfnamefont{T.}~\bibnamefont{Arakane}},
  \bibinfo{author}{\bibfnamefont{T.}~\bibnamefont{Sato}},
  \bibinfo{author}{\bibfnamefont{S.}~\bibnamefont{Souma}},
  \bibinfo{author}{\bibfnamefont{K.}~\bibnamefont{Kosaka}},
  \bibinfo{author}{\bibfnamefont{K.}~\bibnamefont{Nakayama}},
  \bibinfo{author}{\bibfnamefont{M.}~\bibnamefont{Komatsu}},
  \bibinfo{author}{\bibfnamefont{T.}~\bibnamefont{Takahashi}},
  \bibinfo{author}{\bibfnamefont{Z.}~\bibnamefont{Ren}},
  \bibinfo{author}{\bibfnamefont{K.}~\bibnamefont{Segawa}}, \bibnamefont{and}
  \bibinfo{author}{\bibfnamefont{Y.}~\bibnamefont{Ando}},
  \bibinfo{journal}{Nature Commun.} \textbf{\bibinfo{volume}{3}},
  \bibinfo{pages}{636} (\bibinfo{year}{2012}).

\bibitem[{\citenamefont{Segawa et~al.}(2012)\citenamefont{Segawa, Ren, Sasaki,
  Tsuda, Kuwabata, and Ando}}]{Segawa2012}
\bibinfo{author}{\bibfnamefont{K.}~\bibnamefont{Segawa}},
  \bibinfo{author}{\bibfnamefont{Z.}~\bibnamefont{Ren}},
  \bibinfo{author}{\bibfnamefont{S.}~\bibnamefont{Sasaki}},
  \bibinfo{author}{\bibfnamefont{T.}~\bibnamefont{Tsuda}},
  \bibinfo{author}{\bibfnamefont{S.}~\bibnamefont{Kuwabata}}, \bibnamefont{and}
  \bibinfo{author}{\bibfnamefont{Y.}~\bibnamefont{Ando}},
  \bibinfo{journal}{Phys. Rev. B} \textbf{\bibinfo{volume}{86}},
  \bibinfo{pages}{075306} (\bibinfo{year}{2012}).

\bibitem[{\citenamefont{Xu et~al.}(2014)\citenamefont{Xu, Miotkowski, Liu,
  Tian, Nam, Alidoust, Hu, Shih, Hasan, and Chen}}]{QHE-Xu2014}
\bibinfo{author}{\bibfnamefont{Y.}~\bibnamefont{Xu}},
  \bibinfo{author}{\bibfnamefont{I.}~\bibnamefont{Miotkowski}},
  \bibinfo{author}{\bibfnamefont{C.}~\bibnamefont{Liu}},
  \bibinfo{author}{\bibfnamefont{J.}~\bibnamefont{Tian}},
  \bibinfo{author}{\bibfnamefont{H.}~\bibnamefont{Nam}},
  \bibinfo{author}{\bibfnamefont{N.}~\bibnamefont{Alidoust}},
  \bibinfo{author}{\bibfnamefont{J.}~\bibnamefont{Hu}},
  \bibinfo{author}{\bibfnamefont{C.-K.} \bibnamefont{Shih}},
  \bibinfo{author}{\bibfnamefont{M.~Z.} \bibnamefont{Hasan}}, \bibnamefont{and}
  \bibinfo{author}{\bibfnamefont{Y.~P.} \bibnamefont{Chen}},
  \bibinfo{journal}{Nature Phys.} \textbf{\bibinfo{volume}{10}},
  \bibinfo{pages}{956} (\bibinfo{year}{2014}).

\bibitem[{\citenamefont{Yang et~al.}(2016)\citenamefont{Yang, Ghatak, Taskin,
  Segawa, Ando, Shiraishi, Kanai, Matsumoto, Rosch, and Ando}}]{Yang2016}
\bibinfo{author}{\bibfnamefont{F.}~\bibnamefont{Yang}},
  \bibinfo{author}{\bibfnamefont{S.}~\bibnamefont{Ghatak}},
  \bibinfo{author}{\bibfnamefont{A.~A.} \bibnamefont{Taskin}},
  \bibinfo{author}{\bibfnamefont{K.}~\bibnamefont{Segawa}},
  \bibinfo{author}{\bibfnamefont{Y.}~\bibnamefont{Ando}},
  \bibinfo{author}{\bibfnamefont{M.}~\bibnamefont{Shiraishi}},
  \bibinfo{author}{\bibfnamefont{Y.}~\bibnamefont{Kanai}},
  \bibinfo{author}{\bibfnamefont{K.}~\bibnamefont{Matsumoto}},
  \bibinfo{author}{\bibfnamefont{A.}~\bibnamefont{Rosch}}, \bibnamefont{and}
  \bibinfo{author}{\bibfnamefont{Y.}~\bibnamefont{Ando}},
  \bibinfo{journal}{Phys. Rev. B} \textbf{\bibinfo{volume}{94}},
  \bibinfo{pages}{075304} (\bibinfo{year}{2016}).

\bibitem[{\citenamefont{Ghatak et~al.}(2018)\citenamefont{Ghatak, Breunig,
  Yang, Wang, Taskin, and Ando}}]{Ghatak2018}
\bibinfo{author}{\bibfnamefont{S.}~\bibnamefont{Ghatak}},
  \bibinfo{author}{\bibfnamefont{O.}~\bibnamefont{Breunig}},
  \bibinfo{author}{\bibfnamefont{F.}~\bibnamefont{Yang}},
  \bibinfo{author}{\bibfnamefont{Z.}~\bibnamefont{Wang}},
  \bibinfo{author}{\bibfnamefont{A.~A.} \bibnamefont{Taskin}},
  \bibnamefont{and} \bibinfo{author}{\bibfnamefont{Y.}~\bibnamefont{Ando}},
  \bibinfo{journal}{Nano Lett.} \textbf{\bibinfo{volume}{18}},
  \bibinfo{pages}{5124} (\bibinfo{year}{2018}).

\bibitem[{\citenamefont{Jauregui et~al.}(2018)\citenamefont{Jauregui, Kayyalha,
  Kazakov, Miotkowski, Rokhinson, and Chen}}]{Jauregui2018}
\bibinfo{author}{\bibfnamefont{L.~A.} \bibnamefont{Jauregui}},
  \bibinfo{author}{\bibfnamefont{M.}~\bibnamefont{Kayyalha}},
  \bibinfo{author}{\bibfnamefont{A.}~\bibnamefont{Kazakov}},
  \bibinfo{author}{\bibfnamefont{I.}~\bibnamefont{Miotkowski}},
  \bibinfo{author}{\bibfnamefont{L.~P.} \bibnamefont{Rokhinson}},
  \bibnamefont{and} \bibinfo{author}{\bibfnamefont{Y.~P.} \bibnamefont{Chen}},
  \bibinfo{journal}{Appl. Phys. Lett.} \textbf{\bibinfo{volume}{112}},
  \bibinfo{pages}{093105} (\bibinfo{year}{2018}).

\bibitem[{\citenamefont{Liu et~al.}(2018)\citenamefont{Liu, Besbas, Wang, He,
  Chen, Zhu, Wu, Lee, Wang, Moon et~al.}}]{Lu2018}
\bibinfo{author}{\bibfnamefont{Y.}~\bibnamefont{Liu}},
  \bibinfo{author}{\bibfnamefont{J.}~\bibnamefont{Besbas}},
  \bibinfo{author}{\bibfnamefont{Y.}~\bibnamefont{Wang}},
  \bibinfo{author}{\bibfnamefont{P.}~\bibnamefont{He}},
  \bibinfo{author}{\bibfnamefont{M.}~\bibnamefont{Chen}},
  \bibinfo{author}{\bibfnamefont{D.}~\bibnamefont{Zhu}},
  \bibinfo{author}{\bibfnamefont{Y.}~\bibnamefont{Wu}},
  \bibinfo{author}{\bibfnamefont{J.~M.} \bibnamefont{Lee}},
  \bibinfo{author}{\bibfnamefont{L.}~\bibnamefont{Wang}},
  \bibinfo{author}{\bibfnamefont{J.}~\bibnamefont{Moon}}, \bibnamefont{et~al.},
  \bibinfo{journal}{Nature Commun.} \textbf{\bibinfo{volume}{9}},
  \bibinfo{pages}{2492} (\bibinfo{year}{2018}).

\bibitem[{\citenamefont{Efthimiopoulos
  et~al.}(2013)\citenamefont{Efthimiopoulos, Zhang, Kucway, Park, Ewing, and
  Wang}}]{Efthimiopoulos2013}
\bibinfo{author}{\bibfnamefont{I.}~\bibnamefont{Efthimiopoulos}},
  \bibinfo{author}{\bibfnamefont{J.}~\bibnamefont{Zhang}},
  \bibinfo{author}{\bibfnamefont{M.}~\bibnamefont{Kucway}},
  \bibinfo{author}{\bibfnamefont{C.}~\bibnamefont{Park}},
  \bibinfo{author}{\bibfnamefont{R.~C.} \bibnamefont{Ewing}}, \bibnamefont{and}
  \bibinfo{author}{\bibfnamefont{Y.}~\bibnamefont{Wang}},
  \bibinfo{journal}{Sci. Rep.} \textbf{\bibinfo{volume}{3}},
  \bibinfo{pages}{2665} (\bibinfo{year}{2013}).

\bibitem[{\citenamefont{Richter et~al.}(1977)\citenamefont{Richter,
  K{\"{o}}hler, and Becker}}]{Richter1977}
\bibinfo{author}{\bibfnamefont{W.}~\bibnamefont{Richter}},
  \bibinfo{author}{\bibfnamefont{H.}~\bibnamefont{K{\"{o}}hler}},
  \bibnamefont{and} \bibinfo{author}{\bibfnamefont{C.~R.}
  \bibnamefont{Becker}}, \bibinfo{journal}{Phys. Status Solidi}
  \textbf{\bibinfo{volume}{84}}, \bibinfo{pages}{619} (\bibinfo{year}{1977}).

\bibitem[{\citenamefont{Akrap et~al.}(2012)\citenamefont{Akrap, Tran, Ubaldini,
  Teyssier, Giannini, Van Der~Marel, Lerch, and Homes}}]{Akrap2012}
\bibinfo{author}{\bibfnamefont{A.}~\bibnamefont{Akrap}},
  \bibinfo{author}{\bibfnamefont{M.}~\bibnamefont{Tran}},
  \bibinfo{author}{\bibfnamefont{A.}~\bibnamefont{Ubaldini}},
  \bibinfo{author}{\bibfnamefont{J.}~\bibnamefont{Teyssier}},
  \bibinfo{author}{\bibfnamefont{E.}~\bibnamefont{Giannini}},
  \bibinfo{author}{\bibfnamefont{D.}~\bibnamefont{Van Der~Marel}},
  \bibinfo{author}{\bibfnamefont{P.}~\bibnamefont{Lerch}}, \bibnamefont{and}
  \bibinfo{author}{\bibfnamefont{C.}~\bibnamefont{Homes}},
  \bibinfo{journal}{Phys. Rev. B} \textbf{\bibinfo{volume}{86}},
  \bibinfo{pages}{235207} (\bibinfo{year}{2012}).

\bibitem[{\citenamefont{Borgwardt et~al.}(2016)\citenamefont{Borgwardt, Lux,
  Vergara, Wang, Taskin, Segawa, {Van Loosdrecht}, Ando, Rosch, and
  Gr\"uninger}}]{Borgwardt2016}
\bibinfo{author}{\bibfnamefont{N.}~\bibnamefont{Borgwardt}},
  \bibinfo{author}{\bibfnamefont{J.}~\bibnamefont{Lux}},
  \bibinfo{author}{\bibfnamefont{I.}~\bibnamefont{Vergara}},
  \bibinfo{author}{\bibfnamefont{Z.}~\bibnamefont{Wang}},
  \bibinfo{author}{\bibfnamefont{A.~A.} \bibnamefont{Taskin}},
  \bibinfo{author}{\bibfnamefont{K.}~\bibnamefont{Segawa}},
  \bibinfo{author}{\bibfnamefont{P.~H.~M.} \bibnamefont{{Van Loosdrecht}}},
  \bibinfo{author}{\bibfnamefont{Y.}~\bibnamefont{Ando}},
  \bibinfo{author}{\bibfnamefont{A.}~\bibnamefont{Rosch}}, \bibnamefont{and}
  \bibinfo{author}{\bibfnamefont{M.}~\bibnamefont{Gr\"uninger}},
  \bibinfo{journal}{Phys. Rev. B} \textbf{\bibinfo{volume}{93}},
  \bibinfo{pages}{245149} (\bibinfo{year}{2016}).

\bibitem[{\citenamefont{Tian et~al.}(2016)\citenamefont{Tian, Osterhoudt, Jia,
  Cava, and Burch}}]{Tian2016}
\bibinfo{author}{\bibfnamefont{Y.}~\bibnamefont{Tian}},
  \bibinfo{author}{\bibfnamefont{G.~B.} \bibnamefont{Osterhoudt}},
  \bibinfo{author}{\bibfnamefont{S.}~\bibnamefont{Jia}},
  \bibinfo{author}{\bibfnamefont{R.~J.} \bibnamefont{Cava}}, \bibnamefont{and}
  \bibinfo{author}{\bibfnamefont{K.~S.} \bibnamefont{Burch}},
  \bibinfo{journal}{Appl. Phys. Lett.} \textbf{\bibinfo{volume}{108}},
  \bibinfo{pages}{041911} (\bibinfo{year}{2016}).

\bibitem[{\citenamefont{Lee et~al.}(2016)\citenamefont{Lee, Cheng, Weng, Chen,
  Tsuei, Yu, Chou, Chang, Tu, Yang et~al.}}]{Lee}
\bibinfo{author}{\bibfnamefont{C.-K.} \bibnamefont{Lee}},
  \bibinfo{author}{\bibfnamefont{C.-M.} \bibnamefont{Cheng}},
  \bibinfo{author}{\bibfnamefont{S.-C.} \bibnamefont{Weng}},
  \bibinfo{author}{\bibfnamefont{W.-C.} \bibnamefont{Chen}},
  \bibinfo{author}{\bibfnamefont{K.-D.} \bibnamefont{Tsuei}},
  \bibinfo{author}{\bibfnamefont{S.-H.} \bibnamefont{Yu}},
  \bibinfo{author}{\bibfnamefont{M.~M.-C.} \bibnamefont{Chou}},
  \bibinfo{author}{\bibfnamefont{C.-W.} \bibnamefont{Chang}},
  \bibinfo{author}{\bibfnamefont{L.-W.} \bibnamefont{Tu}},
  \bibinfo{author}{\bibfnamefont{H.-D.} \bibnamefont{Yang}},
  \bibnamefont{et~al.}, \bibinfo{journal}{Sci. Rep.}
  \textbf{\bibinfo{volume}{6}}, \bibinfo{pages}{36538} (\bibinfo{year}{2016}).

\bibitem[{\citenamefont{Hulliger and F.}(1976)}]{hulliger_lévy}
\bibinfo{author}{\bibfnamefont{F.}~\bibnamefont{Hulliger}} \bibnamefont{and}
  \bibinfo{author}{\bibfnamefont{L.}~\bibnamefont{F.}},
  \emph{\bibinfo{title}{Structural chemistry of layer-type phases}}
  (\bibinfo{publisher}{D. Reidel}, \bibinfo{year}{1976}).

\bibitem[{\citenamefont{Togo and Tanaka}(2015)}]{phonopy}
\bibinfo{author}{\bibfnamefont{A.}~\bibnamefont{Togo}} \bibnamefont{and}
  \bibinfo{author}{\bibfnamefont{I.}~\bibnamefont{Tanaka}},
  \bibinfo{journal}{Scr. Mater.} \textbf{\bibinfo{volume}{108}},
  \bibinfo{pages}{1} (\bibinfo{year}{2015}).

\bibitem[{\citenamefont{Perdew et~al.}(1996)\citenamefont{Perdew, Burke, and
  Ernzerhof}}]{PhysRevLett.77.3865}
\bibinfo{author}{\bibfnamefont{J.~P.} \bibnamefont{Perdew}},
  \bibinfo{author}{\bibfnamefont{K.}~\bibnamefont{Burke}}, \bibnamefont{and}
  \bibinfo{author}{\bibfnamefont{M.}~\bibnamefont{Ernzerhof}},
  \bibinfo{journal}{Phys. Rev. Lett.} \textbf{\bibinfo{volume}{77}},
  \bibinfo{pages}{3865} (\bibinfo{year}{1996}).

\bibitem[{\citenamefont{Perdew et~al.}(1997)\citenamefont{Perdew, Burke, and
  Ernzerhof}}]{PhysRevLett.78.1396}
\bibinfo{author}{\bibfnamefont{J.~P.} \bibnamefont{Perdew}},
  \bibinfo{author}{\bibfnamefont{K.}~\bibnamefont{Burke}}, \bibnamefont{and}
  \bibinfo{author}{\bibfnamefont{M.}~\bibnamefont{Ernzerhof}},
  \bibinfo{journal}{Phys. Rev. Lett.} \textbf{\bibinfo{volume}{78}},
  \bibinfo{pages}{1396} (\bibinfo{year}{1997}).

\bibitem[{\citenamefont{Larson et~al.}(2000)\citenamefont{Larson, Mahanti, and
  Kanatzidis}}]{PhysRevB.61.8162}
\bibinfo{author}{\bibfnamefont{P.}~\bibnamefont{Larson}},
  \bibinfo{author}{\bibfnamefont{S.~D.} \bibnamefont{Mahanti}},
  \bibnamefont{and} \bibinfo{author}{\bibfnamefont{M.~G.}
  \bibnamefont{Kanatzidis}}, \bibinfo{journal}{Phys. Rev. B}
  \textbf{\bibinfo{volume}{61}}, \bibinfo{pages}{8162} (\bibinfo{year}{2000}).

\bibitem[{\citenamefont{Kresse and Hafner}(1993)}]{PhysRevB.47.558}
\bibinfo{author}{\bibfnamefont{G.}~\bibnamefont{Kresse}} \bibnamefont{and}
  \bibinfo{author}{\bibfnamefont{J.}~\bibnamefont{Hafner}},
  \bibinfo{journal}{Phys. Rev. B} \textbf{\bibinfo{volume}{47}},
  \bibinfo{pages}{558} (\bibinfo{year}{1993}).

\bibitem[{\citenamefont{Kresse and Hafner}(1994)}]{PhysRevB.49.14251}
\bibinfo{author}{\bibfnamefont{G.}~\bibnamefont{Kresse}} \bibnamefont{and}
  \bibinfo{author}{\bibfnamefont{J.}~\bibnamefont{Hafner}},
  \bibinfo{journal}{Phys. Rev. B} \textbf{\bibinfo{volume}{49}},
  \bibinfo{pages}{14251} (\bibinfo{year}{1994}).

\bibitem[{\citenamefont{Kresse and
  Furthm\"uller}(1996{\natexlab{a}})}]{KRESSE199615}
\bibinfo{author}{\bibfnamefont{G.}~\bibnamefont{Kresse}} \bibnamefont{and}
  \bibinfo{author}{\bibfnamefont{J.}~\bibnamefont{Furthm\"uller}},
  \bibinfo{journal}{Comput. Mater. Sci.} \textbf{\bibinfo{volume}{6}},
  \bibinfo{pages}{15 } (\bibinfo{year}{1996}{\natexlab{a}}).

\bibitem[{\citenamefont{Kresse and
  Furthm\"uller}(1996{\natexlab{b}})}]{PhysRevB.54.11169}
\bibinfo{author}{\bibfnamefont{G.}~\bibnamefont{Kresse}} \bibnamefont{and}
  \bibinfo{author}{\bibfnamefont{J.}~\bibnamefont{Furthm\"uller}},
  \bibinfo{journal}{Phys. Rev. B} \textbf{\bibinfo{volume}{54}},
  \bibinfo{pages}{11169} (\bibinfo{year}{1996}{\natexlab{b}}).

\end{thebibliography}
	
\end{document}